\documentclass[11pt]{article}
\usepackage{geometry}
\usepackage[utf8]{inputenc}
\usepackage[T1]{fontenc}
\usepackage{authblk}
\usepackage{amsmath,amssymb,amsthm,bbold}
\usepackage{graphicx}
\usepackage{hyperref}
\usepackage[super,sort&compress]{natbib}
\title{Quantum Fuzzy Orbits\footnote{This manuscript has been conceived as a rectification to the discussion initially published by the author in Ref.\cite{fuentes20}. Additionally, while the original conceptual hierarchy and results are maintained, the narrative structure has been tactfully rearranged to enhance readability.}}
\author{Jes\'us Fuentes}
\date{}
\begin{document}
\maketitle
%	////////////////////////////////////////////////////////////////////
\begin{abstract}

This manuscript surveys quantum operations under the influence of harmonic magnetic fields subject to time variations. The author scrutinises the dynamic interplay of these fields and canonical variables, leading to effects such as squeezing, looping effects, and parametric resonances. The time evolution of observables tethered to a time-dependent quadratic Hamiltonian is examined. The results of this inquiry offer  contributions to the quantum control theory.

\end{abstract}
%	////////////////////////////////////////////////////////////////////
%
%	~
%
%	////////////////////////////////////////////////////////////////////
\section{Introduction}

In the world of microscale manipulation, precise control over the motion of minuscule entities, or qubits, is key. This process typically employs oscillating external fields to guide a qubit within a confined space, inducing specific movements or steering it toward a pre-set configuration\cite{ramakrishna,ramakrishna02,schirmer,haroche,nogues}. However, real-world factors, including environmental noise and electric shocks, pose significant challenges, necessitating the development of techniques to circumvent these hurdles and achieve desired effects in a practical manner.

One approach to tackle this issue involves using reasonable approximations in the control schemes. For example, successful results have been obtained using magnetic nuclear resonance techniques that neglect certain intrinsic inhomogeneities. Furthermore, some methods suppress unwanted effects if the pulses of electrical forces are solely time-dependent\cite{schirmer}. Yet, when a particle is not strictly confined, these techniques may falter. This is where larger cavities come into play, accommodating an electron's free propagation and offering a way to guide its motion precisely, making such techniques useful in applications like particle accelerators\cite{paul,hu,fluhmann,ale2,blais,gessner}.

As the field continues to evolve, new techniques are being considered. For instance, there is increasing interest in the realm of squeezed states\cite{burd}, that can produce efficient non-demolition measurements\cite{thorne2,thorne}, despite their reliance on approximation. We focus on exploring an alternative strategy, particularly looking into time-dependent magnetic fields modulated by control currents. This approach potentially allows us to design magnetic field pulses that guide a charged particle's evolution over time into different types of motion\cite{emma,mielnik77,mielnik86,fernandez94,chen04,emma,harel,viola,mancini,casta,asorey}. Though simplified for practical implementation, this exploration aims to offer insights into the challenges of micro-manipulation.

In agreement with causality, it must be noted that the adjustments in field intensity owing to time-varying currents are not instantly discernible throughout the laboratory. Instead, these effects manifest at different rates, contingent upon the laboratory's dimensions and the signal velocities. From a precise standpoint, one should incorporate the delayed field effects into this problem's treatment. However, for the sake of simplicity, we have limited our analysis to the non-relativistic regime, thus overlooking the time delay in field propagation. This simplification warrants significant consideration for practical implementation, and we intend to rigorously examine the resultant consequences.

%	////////////////////////////////////////////////////////////////////
%
%	~
%
%	////////////////////////////////////////////////////////////////////
\section{Classical-Quantum Duality}

To ensure a self-contained discussion, we begin by revisiting the fundamentals as established by Johnson \cite{johnson}. We consider a scenario involving a spinless particle, characterised by a constant mass $m$ and charge $e$, set in motion within a spatially uniform magnetic field, denoted by $\mathbf{B}(\mathbf{X}) \in \mathbb{R}^3$. For analytical convenience, we align this field along the $Oz$ direction. The origin of this field can be attributed to a certain class of vector potential, represented by $\mathbf{A}(\mathbf{X})$, which adheres to the symmetric form given by:
\begin{equation}\label{vector}
\mathbf{A}(\mathbf{X})=\frac{1}{2}\mathbf{B}(\mathbf{X})\times\mathbf{X}.
\end{equation}
It is important to highlight that, while we may specify $\mathbf{B}(\mathbf{X})$, this gauge condition does not lead to a unique determination of $\mathbf{A}(\mathbf{X})$.

Upon establishing these preliminaries, we progress to a scenario characterised by a time-independent, quadratic Hamiltonian:
\begin{equation}\label{static}
H=\frac{m}{2}\mathbf{V}^2.
\end{equation}
In this expression, $\mathbf{V}$ denotes the velocity operator of the particle, given by $\frac{1}{m}\left[\mathbf{P} - \frac{e}{c}\mathbf{A}(\mathbf{X})\right]$, and $\mathbf{P}$ signifies the momentum operator.

The observables $\mathbf{X}$ and $\mathbf{P}$ adhere to standard commutation relations $[\mathbf{X}_j,\mathbf{P}_k]=i\hbar\delta_{jk}$, $[\mathbf{X}_j,\mathbf{X}_k]=0$ and $[\mathbf{P}_j,\mathbf{P}_k]=0$. The components of operator $\mathbf{V}$ comply with 
\[
[\mathbf{V}_j,\mathbf{V}_k]=\frac{ie\hbar}{m^2c}\epsilon_{jkl}\mathbf{B}_l(\mathbf{X}),
\]
where $\epsilon_{jkl}$ equals 1 (-1) if $j,k,l$ represents an even (odd) permutation of 1,2,3, and is zero otherwise. Furthermore, commutation relations between components of $\mathbf{X}$ and $\mathbf{V}$ are easily deduced, as $\mathbf{A}(\mathbf{X})$ commutes with $\mathbf{X}_j$, leading to $[\mathbf{X}_j,\mathbf{V}_k]=\frac{i}{m}\hbar\delta_{jk}$.

Certain aspects of our discussion warrant particular attention. Unlike the classical analogue where a particle's trajectory can be precisely localised, here, the quantum particle describes a trajectory whose centre $(\bar{X},\bar{Y})$ is best described as a fuzzy point; its precise location remains elusive. The latter is not only due to the non-commuting nature of the position operators, $[\bar{X},\bar{Y}]\neq0$, in fact  this remains true regardless of whether these operators commute or not given the geometric origin remains devoid of physical presence \cite{connes,bellisard}. For instance, consider the gauge \eqref{vector}. The Hamiltonian then splits into two components $H=H_\perp + H_\parallel$: the transverse motion (our primary interest) within the $xOy$ plane and the parallel motion along the $Oz$ axis. The components of the velocity operator $\mathbf{V}$ can be represented as:
\begin{equation}
\begin{split}
V_x&=\frac{P_x}{m} + \frac{\omega_c}{2}Y \\
V_y&=\frac{P_y}{m} - \frac{\omega_c}{2}X \\
V_z&=\frac{P_z}{m},
\end{split}
\end{equation}
where $\omega_c=\frac{e}{mc}B$ is the cyclotron frequency and $B=|\mathbf{B}(\mathbf{X})|$. Consequently, the rotation centre can be envisaged as a fuzzy point, characterised by coordinates:
\begin{equation}\label{fuzzy}
\bar{X}=X+\frac{1}{\omega_c}V_y, \quad \bar{Y}=Y-\frac{1}{\omega_c}V_x,
\end{equation}
which satisfy the commutation relation $[\bar{X},\bar{Y}]=-\frac{i\hbar}{m\omega_c}$. Moreover, should we define
\[
\rho^2=(X-\bar{X})^2+(Y-\bar{Y})^2,
\]
then $\pi\rho^2=\frac{2\pi}{m\omega_c^2}H$ can be interpreted as a fuzzy surface traversed by the transverse orbits. Consequently, while $\pi\rho^2$ remains a conserved quantity, its evolution cannot be continuous \cite{vagner} as it is quantised, a feature with significant implications for loop quantum gravity \cite{ashtekar,rovelli}, for instance.

The appearance of these fuzzy centres extends beyond the realm of quantum Hall-effect models \cite{chalopin} under the influence of a steady magnetic field. They also manifest in time-variant contexts. In our exploration, we shall investigate the role of time-dependent magnetic interactions and their contribution to the non-commutative characteristics of both closed and open semiclassical trajectories.

Simultaneously, we establish a time-dependent variation of Hamiltonian \eqref{static}. Essentially, it can be constructed similarly, but with the consideration of a controlled current $I(t)$ that modulates $\mathbf{A}(\mathbf{X})$ and $\mathbf{B}(\mathbf{X})$. Consequently, their time-dependent equivalents are represented as $\mathbf{A}(t,\mathbf{X}) = I(t)\mathbf{A}(\mathbf{X})$ and $\mathbf{B}(t,\mathbf{X}) = I(t)\mathbf{B}(\mathbf{X})$. Such characterisations imply that the time-dependent  version of \eqref{static} can be expressed as follows \cite{landovitz}:
\begin{equation}\label{hamiltonian}
\begin{split}
H(t) &=\frac{m}{2}\mathbf{V}^2(t) \\
&=\frac{1}{2m}\left[\mathbf{P}-\frac{e}{c}\mathbf{A}(t,\mathbf{X})\right]^2 \\
&= \underbrace{\frac{1}{2m}\left[P_x^2 + P_y^2 + \left(\frac{eB(t)}{2c}\right)^2(X^2+Y^2)\right]}_{H_\text{osc}(t)} - \underbrace{\vphantom{\left[\left(\frac{eB(t)}{2c}\right)^2\right]}\frac{eB(t)}{2mc} L_z}_{H_\text{rot}(t)} + \underbrace{\vphantom{\left[\left(\frac{eB(t)}{2c}\right)^2\right]}\frac{1}{2m}P_z^2}_{H_\parallel}.
\end{split}
\end{equation}

Because $H_\parallel$ yields the widely recognised free particle eigenfunction $\frac{1}{\sqrt{2\pi\hbar}}e^{ip_z z /\hbar}$, we will isolate this element from our analysis. Instead, we focus on $H_\perp(t)=H_\text{osc}(t) + H_\text{rot}(t)$, wherein the oscillatory term $H_\text{osc}(t)$ mirrors a two-dimensional harmonic oscillator, and the readily integrable term $H_\text{rot}(t)$ denotes the rotations effected by $L_z$---the component of the angular momentum oriented along $Oz$---a conserved quantity. This depiction is reminiscent of a cylinder with its symmetry axis parallel to $Oz$ and endowed with a homogeneous current density across its surface.

To advance, it proves beneficial to express the perpendicular terms in \eqref{hamiltonian} in dimensionless variables. Here, we represent the time scale or period of operation as $T=\frac{2\pi}{\omega}$. Our newly introduced variables are then defined as:
\begin{equation}
\label{variables}
t\to \frac{t}{T},\qquad \mathbf{P}\to\sqrt{\frac{T}{\hbar m}}\mathbf{P},\qquad \mathbf{X}\to\sqrt{\frac{m}{\hbar T}}\mathbf{X},
\end{equation}
As a result, the intensity of the oscillatory field is denoted as:
\begin{equation}
\label{beta}
\beta(t)=\frac{eTB(t)}{2mc}.
\end{equation}

The function $\beta(t)$ serves as the critical touchstone for our study. We aim to implement a continuous, real function that resolves the underlying evolution problem. Special attention must be paid to this task; should $\beta(t)$ be bounded and piecewise continuous, the evolution of $\mathbf{X}$ and $\mathbf{P}$ will follow suit. However, this is not applicable to the components of the kinetic momentum in the event of abrupt jumps, interpretable as electric shocks\cite{ale,fan}, characterised by $\mathbf{E}=-\frac{e}{c} \frac{\partial}{\partial t}\mathbf{A}(t,\mathbf{X})$.

The substitution of \eqref{variables} and \eqref{beta} into the perpendicular terms of Hamiltonian \eqref{hamiltonian} leads to a simplified, dimensionless form:
\begin{equation}\label{hd}
H_\perp(t)=\underbrace{\frac{1}{2}\left(\mathbf{P}^2 + \beta^2(t)\mathbf{X}^2\right)}_{H_\text{osc}(t)} - \underbrace{\vphantom{\frac{1}{2}}\beta(t)L_z}_{H_\text{rot}(t)},
\end{equation}
In the particular case of $B(t)$ oscillating periodically (typical Floquet problem) at frequency $\omega$, the substitution $H_\perp(t\omega)\to\frac{H_\perp(t)}{\hbar\omega}$ is also represented in dimensionless terms. It is crucial to note that in this scenario, the stability regions defined by the original Hamiltonian will no longer apply to the dimensionless Hamiltonian.

Although our discussion is primarily concerned with magnetic operations, it is noteworthy that the Hamiltonian \eqref{hd}, despite its elemental structure, can also articulate the evolution of charged particles within the confines of hyperbolically shaped Paul traps \cite{paul} with a designated radius $r_0$. Under such circumstances, the time-dependent electric potentials assume forms such as 
\[
\Phi(t,\mathbf{X})=\frac{e\phi(t)}{2r_0^2}\left(X^2+Y^2-2Z^2\right)
\] 
or 
\[
\Phi(t,\mathbf{X})=\frac{e\phi(t)}{2r_0^2}\left(X^2-Y^2\right).
\] 
These models define a modulated intensity $\beta(t)=\frac{eT^2\phi(t)}{mr_0^2}$.

Interestingly, one discerns that the evolution problem can be tackled in a manner akin to the magnetic issue. In both instances, the resulting evolution matrices $u(t,t_0)$ are identical for classical and quantum dynamics. Furthermore, this close relationship between electric and magnetic problems underlines the versatility of our treatment and emphasises the broad applicability of the Hamiltonian \eqref{hd}, thus enhancing its relevance beyond the immediate context of magnetic operations.

In the subsequent sections, we shall address the evolution problem delineated by \eqref{hd}. We shall initially focus on the previously studied\cite{ale,ale2} biharmonic amplitudes $\beta(t)$, in particular, reassessing the stability map inclusive of the magnetic control operations. This initial stage is our basis to generalise the biharmonic approach, focusing on polyharmonic amplitudes $\beta(t)$, which simultaneously function as exact solutions. It is crucial to underscore that the comprehensive classification of fuzzy orbits within the polyharmonic case remains a compelling open question in our study.

%	////////////////////////////////////////////////////////////////////
%
%	~
%
%	////////////////////////////////////////////////////////////////////
\section{Evolution Loops}

Before we proceed to solve the evolution problem, as represented by \eqref{hd}, we need to revisit the Heisenberg programme. We initiate our discussion with the operator equations that the unitary evolution operators $U(t,t_0)$ satisfy:
\begin{equation}\label{evolution}
\frac{d}{dt}U(t,t_0)=-iH(t)U(t,t_0), \quad \frac{d}{dt_0}U(t,t_0)=iU(t,t_0)H(t_0), \quad U(t_0,t_0)=\mathbb{1},
\end{equation}

For the sake of convenience, and without any loss of generality, we opt to set $t_0=0$. This enables us to introduce the abbreviations $u(t)=u(t,0)$ and $U(t)=U(t,0)$.

We shall use the column-vector $Q$, a $2N$-dimension entity, to represent the set of ordered pairs of observables $\mathbf{X}$ and $\mathbf{P}$ as $Q=(Q_1,\ldots,Q_{2N})^\intercal=((\mathbf{X},\mathbf{P})_1,\ldots,(\mathbf{X},\mathbf{P})_{2N})^\intercal$. In our specific case, $Q=(X,P_x,Y,P_y)^\intercal$. Remember, we are exclusively interested in the motion projected onto the plane $xOy$, as depicted by the Hamiltonian $H_\perp(t)$. Consequently, a set of time-dependent observables $\mathbf{X}(t)$ and $\mathbf{P}(t)$ in the Heisenberg picture will evolve as a linear combination of their initial values $\mathbf{X}$ and $\mathbf{P}$, adhering to the rule:
\begin{equation}\label{heisenberg}
Q(t)=U^\dagger(t)QU(t)=u(t)Q,
\end{equation}

Here, $u(t)$ represents a $2N\times2N$ time evolution matrix that unequivocally determines the evolution operator $U(t)$, and vice versa, but only if $Q$ spans a complete set of observables in a Hilbert space $\mathbb{H}$.

The $j$-th coordinate of the fuzzy centre \eqref{fuzzy} can be written as:
\begin{equation}\label{coordinates}
\bar{X}_j(t)=\bar{u}_{j1}X_1+\bar{u}_{j2}P_1+\cdots+\bar{u}_{j,2N-1}X_{N}+\bar{u}_{j,2N}P_{N},
\end{equation}
where
\begin{equation*}
\quad (X_1,X_2,X_3)=(X,Y,Z), \quad \bar{u}_{jk}=\frac{1}{T}\int_0^\intercal dt\, u_{jk}(t).
\end{equation*}

The direct application of the matrix $u(t)$ to the initial vector $Q$ will yield the entire motion trajectory. Moreover, as $H_\perp(t)$ is quadratic in its variables, the evolution operator $U(t)$ will be identical in the classical and quantum regimes. Consequently, the classical motion trajectories $U(t)$ will have analogous interpretations in the quantum case, where the mean position $\langle \mathbf{X}\rangle$ will evolve according to the Ehrenfest's theorem: $i\hbar\frac{d}{dt}\langle \mathbf{X}\rangle=\langle[\mathbf{X},H_\perp(t)]\rangle$.

It is noteworthy that if two unitary operators $U_1(t)$ and $U_2(t)$ lead to the transformation:
\begin{equation*}
U_1^\dagger(t)QU_1(t)=U_2^\dagger(t)QU_2(t) \quad \text{then} \quad U_2(t)U_1^\dagger(t)Q=QU_2(t)U_1^\dagger(t),
\end{equation*}
and $U_2(t)U_1^\dagger(t)$ commutes simultaneously with any function depending on $\mathbf{X}$ and $\mathbf{P}$. This is particularly significant, as the observables $\mathbf{X}$ and $\mathbf{P}$ generate an irreducible algebra in $L^2(\mathbb{R})$, meaning $U_1(t)U_2^\dagger(t)$ must be a phase factor if these are unitary operators. Hence, we derive $U_1(t)U_2^\dagger(t)=e^{i\varphi}$, or equivalently $U_1(t)=e^{i\varphi}U_2(t)$ where $\varphi$ is a real number. If any two unitary operators differ by a phase factor, they generate identical quantum state transformations, therefore are considered equivalent in this context.

The evolution loop condition\cite{mielnik86} will be satisfied provided the entire set of observables $Q$ return to their initial conditions after a time interval $T$. The analysis is significantly simplified if we assume that \eqref{hd} exhibits a periodic temporal dependence, i.e., $H_\perp(T+t)=H_\perp(t)$, which is indicative of a Floquet Hamiltonian. The periodicity of \eqref{hd} also implies $U(t)=U(T+t)$, concluding that any observable $Q$ in the Heisenberg picture will evolve periodically, even if it does not explicitly depend on time. It follows from Eq. \eqref{heisenberg} that
\[
Q(T+t)=U^\dagger(T+t)QU(T+t)=Q(t),
\]
or for a fixed period of time evolution $Q(T)=U^\dagger(T)QU(T)=Q$, implying that the loop condition holds if $U(T)=e^{i\varphi}\mathbb{1}$.

%	////////////////////////////////////////////////////////////////////
%
%	~
%
%	////////////////////////////////////////////////////////////////////
\section{Biharmonic Fields}

We shall now focus on the analysis of the evolution loop trajectories. These trajectories are part of a system where a solenoid is energised by a surface current density that varies uniformly over time. We will also examine the squeezing effects manifested through biharmonic oscillations of the form\cite{ale}:
\begin{equation}\label{bi}
\beta(t)=\beta_0+\beta_1\sin(\omega_1 t) + \beta_2 \sin(\omega_2 t), \quad \omega_1,\omega_2\in\mathbb{R},
\end{equation}
here, one can engage the harmonic case by setting $\beta_2=0$ or $\omega_2=0$.

The temporal profile of $\beta(t)$ is not confined to a specific form; it can be remarkably arbitrary. However, any viable physical implementation of $\beta(t)$ should ideally steer clear of sudden jumps, such as those delivered by kicked protocols \cite{ale,harel,emma,fan}. These abrupt transitions, in the form of electric delta shocks, may disrupt smooth evolution. Consequently, it might be more beneficial to opt for a sufficiently smooth pulse shaped, for instance, by harmonic functions.

A significant consequence of our cylindrical model is that the commutation relation $[H_\text{osc}(t),H_\text{rot}(t')]=0$ is invariably met, facilitating the division of the evolution operator $U(t)$ into two distinct parts: $U(t)=U_\text{rot}(t)U_\text{osc}(t)$. These two components respectively describe the evolution around $Oz$ and the evolution of the magnetic oscillator $U_\text{osc}(t)$, both satisfying \eqref{evolution}. Consequently, the resulting evolution matrix $u(t)$ evolves from the initial condition $Q$ to $Q_\text{osc}(t)$ to $Q(t)$, directed by the successive application of the oscillatory and rotational evolutions.

The operator $U_\text{rot}(t)$ generates the rotation matrix $u_\text{rot}(t)$, which can be easily integrated as follows:
\begin{equation}\label{urot}
U_\text{rot}(t)=e^{-iL_z\int_0^t dt'\beta(t')},
\end{equation}

The $4\times4$ matrix $u_\text{osc}(t)$ is constructed from $U_\text{osc}(t)$. The integration of the oscillatory part is simplified further as $u_\text{osc}(t)$ can be represented by a single matrix $h(t)$. This matrix enables the simultaneous evolution of each observable pair $Q_x=(X,P_x)^\intercal$ and $Q_y=(Y,P_y)^\intercal$. Therefore,
\begin{equation}\label{uosc}
u_\text{osc}(t)=\begin{pmatrix} h(t)& \\ & h(t)\end{pmatrix}
\end{equation}
simultaneously influences each of the subspaces spanned by $Q_x$ and $Q_y$. For $Q_j$ where $j=x,y$, we have:
\begin{equation}
\frac{dh(t)}{dt}Q_j = iU_\text{osc}^\dagger(t)[H_\text{osc}(t), Q_j]U_\text{osc}(t)=\Lambda(t)U_\text{osc}^\dagger(t)Q_jU_\text{osc}(t)=\Lambda(t)h(t)Q_j,
\end{equation}
where
\begin{equation}
\Lambda(t)=\begin{pmatrix}0&1\\-\beta^2(t)&0\end{pmatrix},
\end{equation}

The conclusion drawn from this is that $h(t)$ simply obeys the differential equation:
\begin{equation}\label{hdiff}
\frac{dh(t)}{dt}=\Lambda(t)h(t), \quad h(0)=\mathbb{1}_{2\times2},
\end{equation}
while the integration of this equation becomes standard for stationary fields, $\beta(t)=\beta_0$, resulting in solutions of the form $e^{\Lambda t}$, we generally prefer to have a computer perform this task.

It is important to note that the determinant of the symplectic matrix $h(t)$ is an integral of the motion, such that $\text{det}(h(t))=1$. This property is essential for classifying the particle's motion and finding exact solutions, as will be demonstrated below.

The matrix $h(t)$ is crucial in the evolution process steered by the class of periodic fields designated as $\beta(T+t)=\beta(t)$. Hence, understanding the dynamical attributes inherent to such matrix becomes an essential pursuit.

Indeed, $h(t)$ carries the distinctive feature of being symplectic, exhibiting two eigenvalues: $\lambda^+$ and $\lambda^-$, conforming to the relation $\lambda^+=\frac{1}{\lambda^-}$. This lays out a unique algebraic structure for $h(t)$, determined entirely by the scalar $\Sigma=\text{tr}(h(t))$. The characteristic polynomial $D(\lambda)=\lambda^2-\Sigma\lambda+1$ and its roots $\lambda^\pm=\frac{1}{2}(\Sigma\pm\sqrt{\Delta})$, where $\Delta=\Sigma^2-4$, further underline this structure. Through $|\Sigma|$, we are then equipped to classify the motion demonstrated into three distinct categories \cite{ale}:

\begin{description}
\item[$|\Sigma|<2$.] In this scenario, $h(t)$ possesses the eigenvalues $\lambda^+=e^{+i\sigma}$ and $\lambda^-=e^{-i\sigma}$, where $\sigma\in[0,\frac{\pi}{2}]$. The motion here is entirely stable, constrained to an oscillatory evolution. This property enables us to extract the annihilation and creation operators, $a^-$ and $a^+$, from the eigenvectors of $h(t)$. In practical applications, quantum operations of this type have proven beneficial in areas such as ion traps.

\item[$|\Sigma|=2$.] Here, the eigenvalues of $h(t)$ simplify to $\pm1$, acting as a boundary between stable and unstable regions. The motion under this field amplitude threshold can induce an effective parametric resonance. This capability lends itself to an array of intriguing applications. For instance, a charged particle can be drawn towards repulsive forces, defying conventional expectations.

\item[$|\Sigma|>2$.] For this category, $h(t)$ bears a pair of real eigenvalues: $\lambda^+=e^{+\sigma}$ and $\lambda^-=e^{-\sigma}$, with $\sigma\in[0,\frac{\pi}{2}]$. This motion type is inherently unstable, leading to a squeezing effect on the annihilation and creation operators, $a^-$ and $a^+$. This leads to an increase in $a^-$ at the expense of $a^+$, or vice versa.
\end{description}

The categorisation of motion contingent upon the value of $|\Sigma|$ equips us with an intuitive framework for comprehending and predicting dynamical patterns under variable field conditions.

Even though the above classification pivots entirely on Heisenberg's evolution of canonical observables, it is worth noting that the parametric resonance region, $|\Sigma|=2$, can be envisaged as an analogue of the parametric amplification. This was initially studied by Mollow and Glauber\cite{mollow} within the context of coherent photon states. An interesting feature here is that trajectory visuals often draw the interest of researchers working on the design of ion traps.

\begin{figure}[htbp]
\begin{center}
\includegraphics[width=14cm]{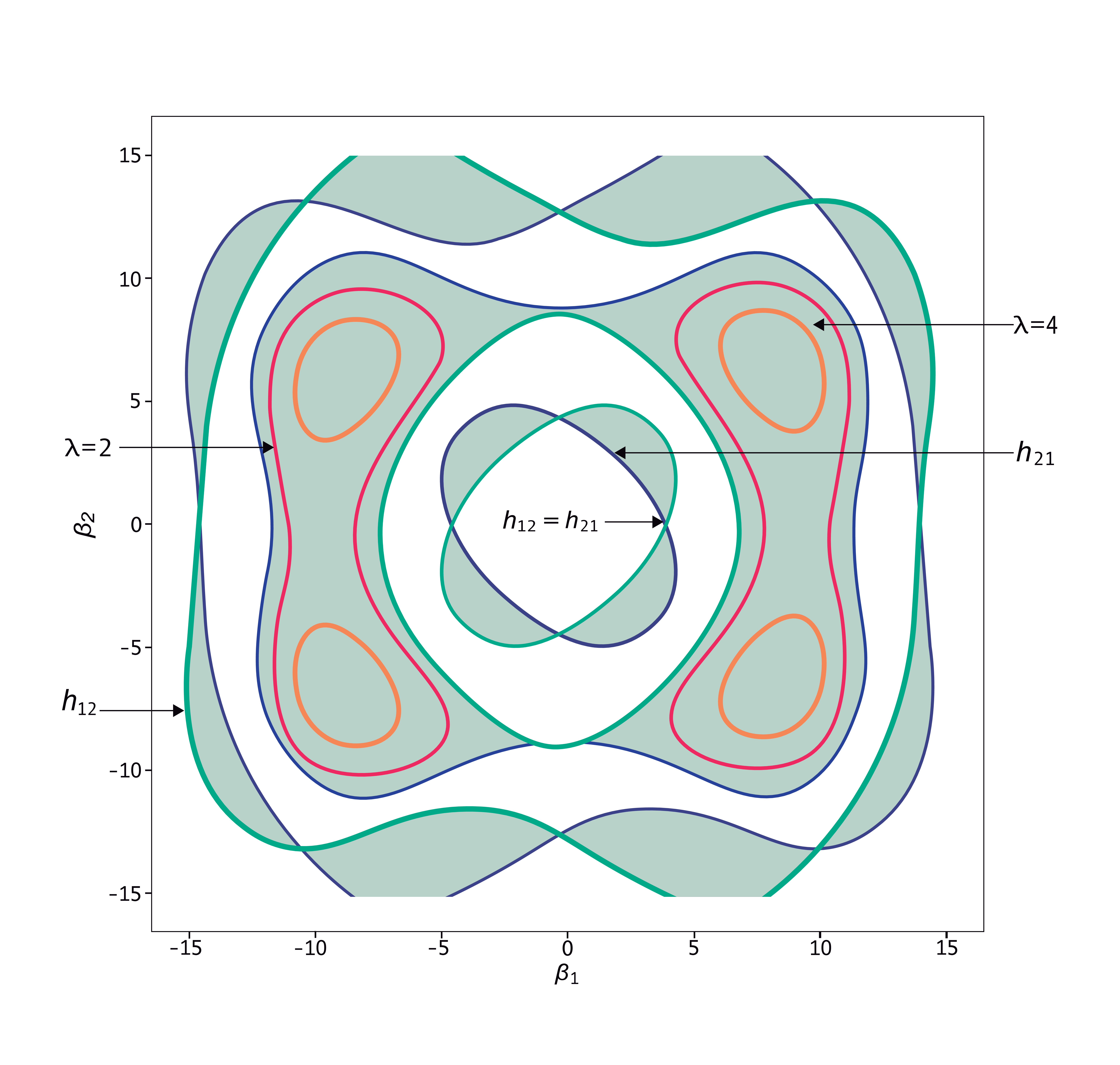}
\caption{Strutt diagram of a biharmonic field \eqref{bi}. The stability regions, $|\Sigma|<2$, correspond to the clear areas, whereas the instability regions, $|\Sigma|>2$, appear in colour. The threshold belts, $|\Sigma|=2$, can be tracked with the aid of the matrix entries $h_{11}$ or $h_{21}$. If both entries are simultaneously equal to zero we have a pure squeezing effect, otherwise, the squeezing operations can be generated by taking the points inside the coloured areas. That is the case of the tracked subregions of amplitudes that would achieve the effects of $h_{11}=\lambda=2$ or $h_{11}=\lambda=4$. A point $(\beta_1,\beta_2)$ lying in the clear areas could be used to produce a class of stable motion, showing a loop effect, such as the required for ion traps.}
\label{f:strutt}
\end{center}
\end{figure}

As an example, a well known ion trap is the radio-frequency, quadrupole device engineered by Paul\cite{paul} in which the oscillating elastic forces $\beta(t)$ are written in terms of Mathieu functions. Indeed, the stability regions for such trap are simply identified from a Strutt map, useful in a variety of time-dependent problems with elastic potentials, e.g. quantum tomography\cite{mancini}.

%	////////////////////////////////////////////////////////////////////
%
%	~
%
%	////////////////////////////////////////////////////////////////////
\subsection{Simple Numerical Explorations}

Turning our focus to the practical applications of this classification, we explore its effectiveness not only in the realm of ion traps, but also its potential within the expansive landscape of quantum control operations. 

To demonstrate this, we have coded a computer routine to numerically integrate \eqref{hdiff} utilising a biharmonic field \eqref{bi} with a fixed $\beta_0=0$. A thorough scanning process is performed to locate the different values of $|\Sigma|$ in the amplitude domain. We have selected $\beta_1,\beta_2\in[-15,15]$ with angular frequencies $\omega_1=2\pi$, $\omega_2=4\pi$ and $t\in[0,1]$. Consequently, we have charted the biharmonic map displayed in Fig. \ref{f:strutt}, wherein the three regions according to $|\Sigma|$ can be found in the amplitude space delineated by ${\beta_1,\beta_2}$. Simply put, based on the values of these amplitudes, a certain type of quantum control operation will be produced.

\begin{figure}[t]
\begin{center}
\includegraphics[width=1.1\textwidth]{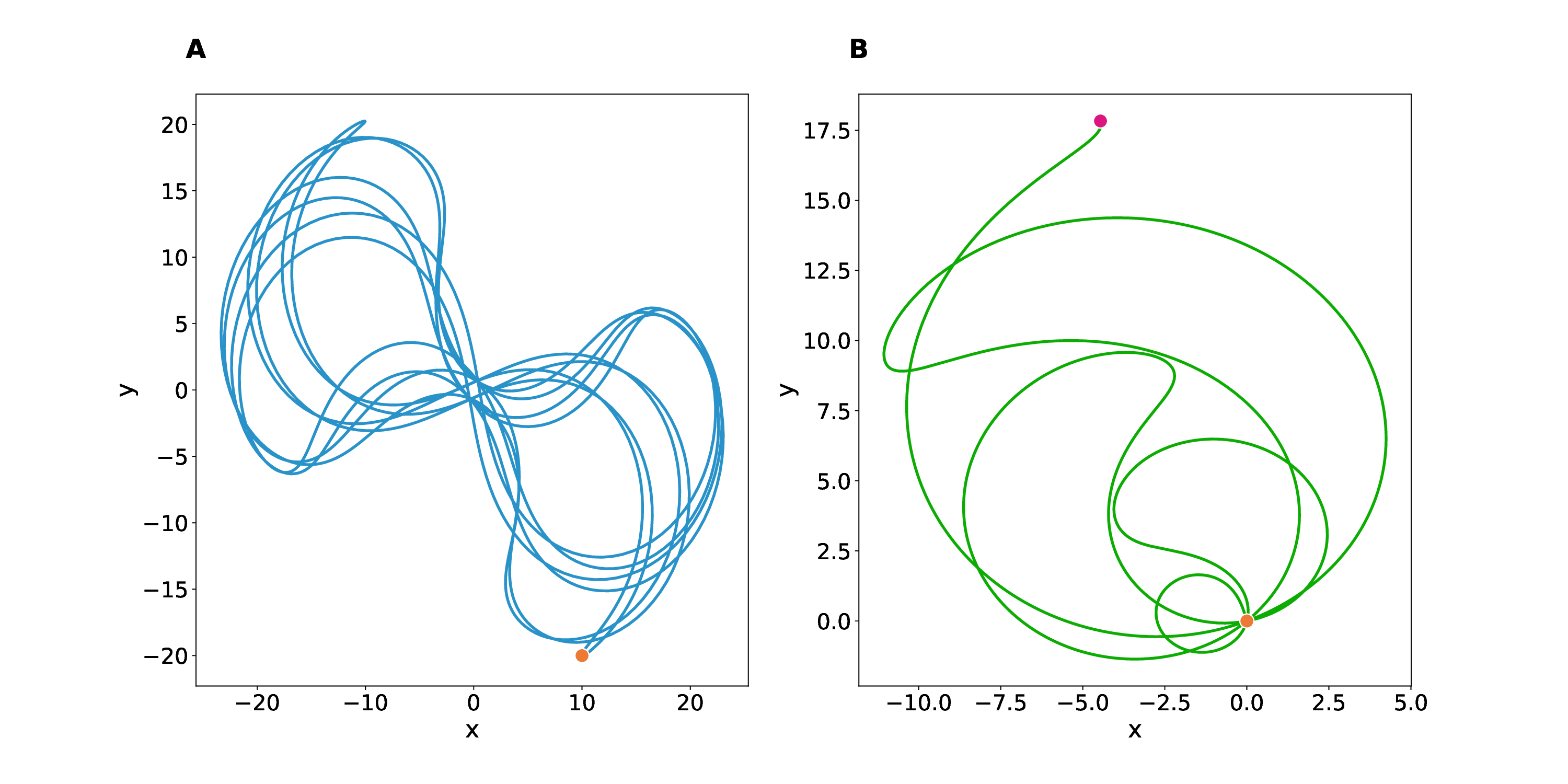}
\caption{{\bf A.} Depiction of a semiclassical closed trajectory in the $xOy$ plane. This trajectory originates from the initial conditions $(x,y)=(10,-20)$ and $(p_x,p_y)=(-5,20)$. The elastic field's corresponding dimensionless amplitudes, $\beta_1=\frac{\pi}{4}$ and $\beta_2=-10$, lie within the stability region, as illustrated in Fig. \ref{f:strutt}. The trajectory culminates in a loop after a period of $T=6$. The orange dot denotes both the initial and final positions of the trajectory. {\bf B.} Representation of a fragmented loop of inverted free evolution. This trajectory starts from the initial conditions $(x,y)=(0,0)$ and $(p_x,p_y)=(-5,20)$. It is associated with the amplitudes $\beta_1=-593/50$ and $\beta_2=-4/10$, which fall precisely on the separatrix between the regions of stability and instability. Control operations executed with these types of amplitudes are completed over a period of $T=2$. The orange and red dots signify the initial and final positions of the trajectory, respectively.}
\label{f:bi}
\end{center}
\end{figure}

Our first task is to examine the stable motion guaranteed by the amplitude set ${\beta_1,\beta_2}$ situated within the clear areas of the map in Fig. \ref{f:strutt}. To shed light on how evolution loops originate, we have integrated \eqref{hdiff} for up to $T=6$ field periods, resulting in the closed trajectory displayed in Fig. \ref{f:bi}-A. The corresponding dimensionless amplitudes for this trajectory are $\beta_1=\frac{\pi}{4}$ and $\beta_2=-10$. In this instance, the particle initiates its evolution with velocities $(p_x,p_y)=(-5,20)$ at the point $(x,y)=(10,-20)$ and concludes at the same point with velocities $(p_x,p_y)=(25/4, -1-\sqrt{2})$. Any evolution loop with symmetry under parity reflection, akin to this one, will exhibit a vanishing fuzzy point $(\bar{X},\bar{Y})$, signifying resilience to the influence of external driven forces. In fact, whenever the elastic field $\beta(t)$ is biharmonic, the potential $\mathbf{A}(t,\mathbf{X})$ will disappear at the beginning $t=0$ and at the end $t=T$ of the process, thereby causing the canonical momenta and the kinetic momenta to coincide. This aids in simplifying the interpretation of the wave packet at these points.

This oscillatory motion is not strictly confined to evolutionary circuits. It can, under certain conditions, fragment into open trajectories. These cases are of particular interest as they incorporate a form of free evolution within the particle's motion history. Moreover, in the Schr\"odinger picture, the free evolution over a time interval $\tau$ corresponds to the operator $e^{-\frac{i\tau}{\hbar}\frac{\mathbf{P}^2}{2}}$, which alters the canonical observables into $X_j\to X_j+\tau P_j$ and $P_j\to P_j$ via an evolution matrix of the form \eqref{uosc}. In this scenario, however, $h(t) \to h(\tau)$ adopts a simplistic structure:
\begin{equation}
h(\tau)=\begin{pmatrix}1&\tau \\ 0&1\end{pmatrix}.
\end{equation}
Subsequently, if $h(\tau)$ depicts a segment of an evolution loop, the rest of the trajectory simply corresponds to the inverse operation $e^{+\frac{i\tau}{\hbar}\frac{\mathbf{P}^2}{2}}$, which is $h(-\tau)$, ensuring that $h(\tau)h(-\tau)=\mathbb{1}$. In essence, the action of $h(-\tau)$ reverses the effects produced by $h(\tau)$ over the set of observables, causing the wave packet to be focused rather than contributing to its dispersion across the operation interval.

Such outcomes are possible when the field $\beta(t)$ encompasses amplitudes ${\beta_1,\beta_2}$ that exist within the threshold region, $|\Sigma|=2$. This scenario paves the way for three kinds of temporal manipulations, contingent upon the effective operation time, $\tau$:
\begin{enumerate}
\item Quickened free evolution, $\tau>T$, where the system ages more rapidly than the actual operation time; 
\item Delayed free evolution, $0<\tau<T$, in which the system undergoes a slowed progression of temporal motion; 
\item Inverted free evolution, $\tau<0$, an effect brought about by the consecutive application of two biharmonic periods $T=2$ provided that $\Sigma=-2$, for which $h_{11}=h_{22}=-1$, and consequently $\tau=-2h_{12}$, with $h_{12}>0$.
\end{enumerate}

Consider, for instance, the separatrix point $(\beta_1,\beta_2)=(-593/50,-4/10)$ on the map in Fig. \ref{f:strutt}. Beginning with the initial conditions $(x,y)=(0,0)$ and $(p_x,p_y)=(-5,20)$, the particle will reach the point $(x,y)=(-447/100,357/20)$ with the same velocities after two periods, landing a distance $2\tau|v|$ behind its initial location, as depicted in Fig. \ref{f:bi}-B.

Finally, we are to examine the squeezing effects afforded by points within the coloured regions of the Strutt map, Fig. \ref{f:strutt}, see also \cite{fuentes23}. Our forthcoming discussion is not aimed at the philosophy of squeezed-states but centres on the source of squeezing operations in relation to the periodic-evolution model \eqref{hd}.

The symplectic structure of the matrix $h(T)$ enables us to streamline the analysis and formulate the squeezing condition in terms of its eigenvalues:
\begin{equation}
\label{condition}
\lambda^+\lambda^-=1,
\end{equation}
which holds true inside the region $|\Sigma|>2$. Nevertheless, the time evolution expressed by \eqref{hdiff} must be calculated within a time span where $\beta^2(t)$ is antisymmetric about the interval centre, or squeezing effects will fail to manifest\cite{wolf}. Depending on the actual value of $\Sigma$, there are two potential squeezing transformations: the purely positive transformation for $\Sigma>2$ and the parity transformation for $\Sigma<2$. The choice between these two alternatives, using the points $(\beta_1,\beta_2)$, hinges on the control aims.

Given the condition \eqref{condition}, it follows that the matrix elements $h_{12}$ and $h_{21}$ equate to zero. Consequently, any canonical observable $Q$ will transform as:
\begin{equation}
\label{squeezing}
Q'\to\lambda Q', \quad Q''\to\frac{1}{\lambda}Q'',
\end{equation}
interpreted as an amplification of $Q'$ compensated by the compression of $Q''$, or conversely. These transformations are encapsulated within the eigenvectors of $h(T)$.

We observe that if a fuzzy point \eqref{fuzzy} does not evaporate, the squeezing operations will register the impact of any external force but in a direction orthogonal to its exertion. Otherwise, the control operations become impervious to external perturbations, such as radiation pollution\cite{vepsalainen}.

Nevertheless, the transformations \eqref{squeezing} concerning eigenvalues $x_j\to\lambda x_j$, $p_j\to\frac{1}{\lambda} p_j$ in intervals $[nT,(n+1)T]$ can only materialise at those separatrix points where $h_{12}=h_{21}=0$ is true. This usually happens at the intersection points of matrix trajectories in the separatrix belt. The squeezing transformations \eqref{squeezing} are viable if the amplitudes $\beta_1,\beta_2$ occupy values inside the unstable regions demarcated by the Strutt map. We have identified some of these points at which pronounced squeezing $\lambda=h_{11}$ is accomplished, especially for $\lambda=2$ and $\lambda=4$. For instance, a biharmonic field with the amplitudes $(\beta_1,\beta_2)=(-103/10,-69/10)$ can reach a squeezing/amplification factor $\lambda=4$.

%	////////////////////////////////////////////////////////////////////
%
%	~
%
%	////////////////////////////////////////////////////////////////////
\section{Polyharmonic Fields: Time Evolution \`a la Mielnik}

The quest to devise quantum control operations, such as those outlined in Eq. \eqref{squeezing}, has led to an assortment of methodological approaches. Among these is the programme of discrete pulses, also known as kicked pulses, which serve to interrupt continuous evolution processes\cite{ale,harel,emma}. However, these operations have proven technically challenging, motivating a shift towards a methodology based on softer evolution operations. A partial solution is offered by replacing the discrete pulses with smooth pulses, for example, biharmonic fields $\beta(t)$. This approach mitigates some of the imperfections inherent in the kicked pulses.

In this modified approach, the operation's design might involve the use of two distinct smooth pulses\cite{fan}. These pulses both reside in the stable region of the Strutt diagram but are chosen such that their product falls within the instability region. As a result, the sequential application of these pulses imparts a certain amount of energy to the particle\cite{grubl}, which should not exceed the energy difference between adjacent energy levels of the particle. However, this methodology does not entirely eliminate the problem of sudden jumps in the composition of quantum control operations. The following discussion, though not definitive, will outline an approach to mitigate this issue.

Mielnik offered a solution to this problem in 2013 by suggesting a method to resolve the initial value problem \eqref{hdiff}, sidestepping the need for adiabatic invariants\cite{mielnik13}. We shall revisit his methodology and seek to derive evolution matrices that allow softer transitions in quantum operations.

In our interpretation of Mielnik's approach, the composition methodology involves considering specific segments of the time evolution dictated by the Hamiltonian \eqref{hd}. The sole distinction this time is the transformation of the elastic fields \eqref{beta} from $\beta^2(t)$ to $\beta(t)$, a shift aimed at facilitating the pursuit of exact solutions. As a consequence, within the oscillation subspace, the matrix $h(t)$ adopts the form of a symplectic matrix of rotations:
\begin{equation}\label{hrot}
h(t)=\begin{pmatrix}\cos(\omega t) & \frac{1}{\omega}\sin(\omega t) \\ -\omega \sin(\omega t) & \cos(\omega t)\end{pmatrix},
\end{equation}

For $\omega t=\frac{n\pi}{2}$, where n is an integer, we achieve squeezing transformations:
\begin{equation}
h\left(\frac{n\pi}{2\omega}\right)=\begin{pmatrix}0&\pm\frac{1}{\omega} \\ \mp \omega&0\end{pmatrix},
\end{equation}
The subsequent application of two matrices of this form results in:
\begin{equation}\label{matrixsq}
h_\text{squeezing} = \begin{pmatrix}0&\pm\frac{1}{\omega_1}\\ \mp\omega_1&0\end{pmatrix}\begin{pmatrix}0&\pm\frac{1}{\omega_2}\\ \mp\omega_2&0\end{pmatrix}=\begin{pmatrix}\lambda&0\\ 0&\frac{1}{\lambda}\end{pmatrix}, \quad \lambda = -\frac{\omega_2}{\omega_1}, \quad \omega_1\neq\omega_2,
\end{equation}
This process delivers the set of operations delineated in \eqref{squeezing}, albeit with the stipulation of two distinct frequencies, $\omega_1$ and $\omega_2$, at two distinct times, $t_1$ and $t_2$. These frequencies and times must satisfy the condition $\omega_1t_1=\omega_2t_2=\frac{\pi}{2}$. However, this method does not entirely remove sudden jumps, for instance, when applied in a void background, at least three jumps occur: $0\to\omega_1\to\omega_2\to0$. The challenge now is to translate this discrete method into a continuous operation.

We have demonstrated that the operations outlined in \eqref{squeezing} are generated by symplectic matrices $h(t)$ with the characteristic $h_{11}=h_{22}=\frac{1}{2}\Sigma$, meaning that $h(t)$ adopts the form of a Toeplitz matrix. While it is attractive to assume that if $\eta$ and $\xi$ are Toeplitz matrices, their anti-commutator algebra $\eta\xi + \xi\eta$ and symmetric products $\eta\xi\eta$ and $\xi\eta\xi$ yield matrices of the same class, this is not always the case. In fact, only the sum or difference (including anti-commutators) of two Toeplitz matrices is guaranteed to be a Toeplitz matrix, while the product of two Toeplitz matrices does not necessarily retain the Toeplitz structure.

The squeezing transformations in \eqref{squeezing} can, however, be constructed from a multitude of symmetric products between symplectic matrices $\eta(t)$ of the form \eqref{hrot}. Each one is attributed to different time subintervals $t_j$ with a definitive elastic field amplitude $\beta(t_j)$, where $j=0,1,2,\ldots$. We can express this in the following form:
\begin{equation}\label{sandwich}
h(t)= \eta(t_k) \cdots \eta(t_1) \eta(t_0) \eta(t_1) \cdots \eta(t_k),
\end{equation}
this still upholds the property $h_{11}=h_{22}=\frac{1}{2}\Sigma$.

To translate this into a continuous analogue, we shall assume that the infinitesimal jumps $dh(t)$ are assembled from the infinitesimal contributions $d\eta(t)=\Lambda(t)dt$. This progression is from the right to the left sides, as shown in \eqref{sandwich}, each reliant on a symmetric field $\beta(t)=\beta(-t)$ around $t=0$. This leads to a matrix differential equation in the expanded interval $[-t,t]$:
\begin{equation}\label{constrain}
\frac{dh(t)}{dt} = \Lambda(t)h(t)+h(t)\Lambda(t), \quad h(0)=\mathbb{1}, \quad \Lambda(t)=\begin{pmatrix}0&1\\-\beta(t)&0\end{pmatrix},
\end{equation}
explicitly, this can be written as:
\begin{equation}
\frac{dh(t)}{dt} = \begin{pmatrix}h_{21}-h_{12}\beta(t) & \Sigma \\
-\Sigma\beta(t) &h_{21}-h_{12}\beta(t)\end{pmatrix}=(h_{21}-h_{12}\beta(t))\mathbb{1}+\Sigma\begin{pmatrix}0&1\\-\beta(t) & 0\end{pmatrix},
\end{equation}
the actual trajectory, as determined by the full evolution process, necessitates the integration of \eqref{hdiff} over an asymmetric interval. This is due to the fact that the nature of \eqref{constrain} serves as a guide, as will be clarified later. Moreover, the anti-commutative structure of \eqref{constrain} frames $\beta(t)$ in terms of a smooth enough, real function $\theta(t)$. Our next objective is to ascertain how to derive an exact solution to this inverse evolution problem.

In pursuit of our goal, first consider that the diagonal elements of $h(t)$ fulfil
\[
\frac{dh_{11}}{dt}=\frac{dh_{22}}{dt}=h_{21}-h_{12}\beta(t). 
\]
Given the initial condition $h_{11}=h_{22}=1$ at $t=0$, we are able to express:
\begin{equation}\label{h11}
h_{11}=h_{22}=\frac{1}{2}\frac{d\theta(t)}{dt},
\end{equation}
from this, it follows consistently that $\text{det}(h(t))=(\frac{1}{2}\frac{d\theta(t)}{dt})^2-h_{21}\theta(t)=1$, which leads to:
\begin{equation}
\label{h21}
h_{21}=\frac{\left[\frac{1}{2}\frac{d\theta(t)}{dt}\right]^2-1}{\theta(t)},
\end{equation}
upon substituting into \eqref{h11} and rearranging the terms, we find $h_{12}\beta(t)=h_{21}-\frac{dh_{11}}{dt}$. Furthermore, since $\theta(t)=h_{12}$, it directly implies $\frac{dh_{11}}{dt}=\frac{1}{2}\frac{d^2\theta(t)}{dt^2}$, which gives us:
\begin{equation}\label{exact}
\beta(t)=-\frac{\frac{d^2\theta(t)}{dt^2}}{2\theta(t)}+\frac{\left[\frac{1}{2}\frac{d\theta(t)}{dt}\right]^2-1}{\theta^2(t)},
\end{equation}
it is immediately apparent that the special case $\theta(t)=\frac{1}{\omega}\sin(2\omega t)$ corresponds to the basic harmonic oscillator $\beta(t)=\omega^2=\text{constant}$

Even without specialised insight on adiabatic invariants\cite{suslov,berry}, $\beta(t)$ as articulated in \eqref{exact} is an exact solution to the inverse evolution problem for $h(t)$ given a function $\theta(t)$. This function can, in principle, be arbitrary but must meet non-trivial conditions at singular points. Notwithstanding, for any asymmetric interval $[t_0,t]$, the dependence of $h(t)$ on $\beta(t)$ must be determined by integrating \eqref{hdiff}, as demonstrated earlier for the case of biharmonic fields.

Certain straightforward correlations between $\beta(t)$ and $\theta(t)$ warrant observation. Any field amplitude defined by \eqref{exact} in a symmetric interval $[-t,t]$, is achieved by a function $\theta(t)$ sufficiently smooth to ensure continuity and differentiability. In particular, at any point $t$ where $\theta(t)=0$, it must hold that $\frac{d\theta(t)}{dt}=\pm 2$. Similarly, if $\theta(t)\neq0$ but $\frac{d\theta(t)}{dt}=0$ then \eqref{constrain} becomes a matrix of squeezing transformations akin to the one in \eqref{matrixsq}. Furthermore, if $\frac{d^3\theta(t)}{dt^3}=0$, it follows that $\frac{d\beta(t)}{dt}=0$.

%	////////////////////////////////////////////////////////////////////
%
%	~
%
%	////////////////////////////////////////////////////////////////////
\subsection{Crafting the Polyharmonic Function $\theta(t)$}

The final component required in our analysis is the construction of $\theta(t)$. This function, though largely arbitrary, invites an empirical selection based upon harmonic functions. Such a choice would foster a smooth control operation, avoiding abrupt leaps in the evolution process. We first look at the elementary case of the polyharmonic function:
\begin{equation}\label{theta}
\theta(t) = \sum_{k=1}^{4} a_{2k-1} \sin(\omega_{2k-1} t),
\end{equation}
given the antisymmetry of this function around $t=0$, the corresponding $\beta(t)$---as per \eqref{exact}---will inherently possess symmetry around the same point. Particularly, if we set the frequencies such that $\omega_{2k-1} = 2k-1$ for $k = 1, \ldots, 4$, we will find a squeezing matrix \eqref{matrixsq} at the endpoints of the symmetric interval $\left[-\frac{\pi}{2},\frac{\pi}{2}\right]$ with $h_{12}=\pm\omega=b$. This necessitates the following initial conditions:
\begin{equation}
\begin{pmatrix}
\theta\left(\frac{\pi}{2}\right)\\
\theta'(0)\\
\theta''\left(\frac{\pi}{2}\right)\\
\theta'''(0)
\end{pmatrix} =
\begin{pmatrix}
1 & -1 & 1 & -1 \\
1 & 3 & 5 & 7 \\
-1 & 9 & -25 & 0 \\
-1 & -27 & -125 & -343
\end{pmatrix}
\begin{pmatrix}
a_1 \\
a_3 \\
a_5 \\
a_7
\end{pmatrix}=
\begin{pmatrix}
b \\
2 \\
-\frac{2}{b} \\
c
\end{pmatrix}
\end{equation}
here, $c$ is a non-zero real parameter, which should be adjusted thoughtfully to enable the desired control operation. This condition set casts light on the intricate relationship between $\beta(t)$ and $\theta(t)$. The $\theta\left(\frac{\pi}{2}\right)=b$ condition regulates the magnitude of the squeezing transformation across the entire trajectory, while $\theta'(0)=2$ ensures non-singularity of $\beta(t)$ at $t=0$.

\begin{figure}[h]
\begin{center}
\includegraphics[width=0.8\textwidth]{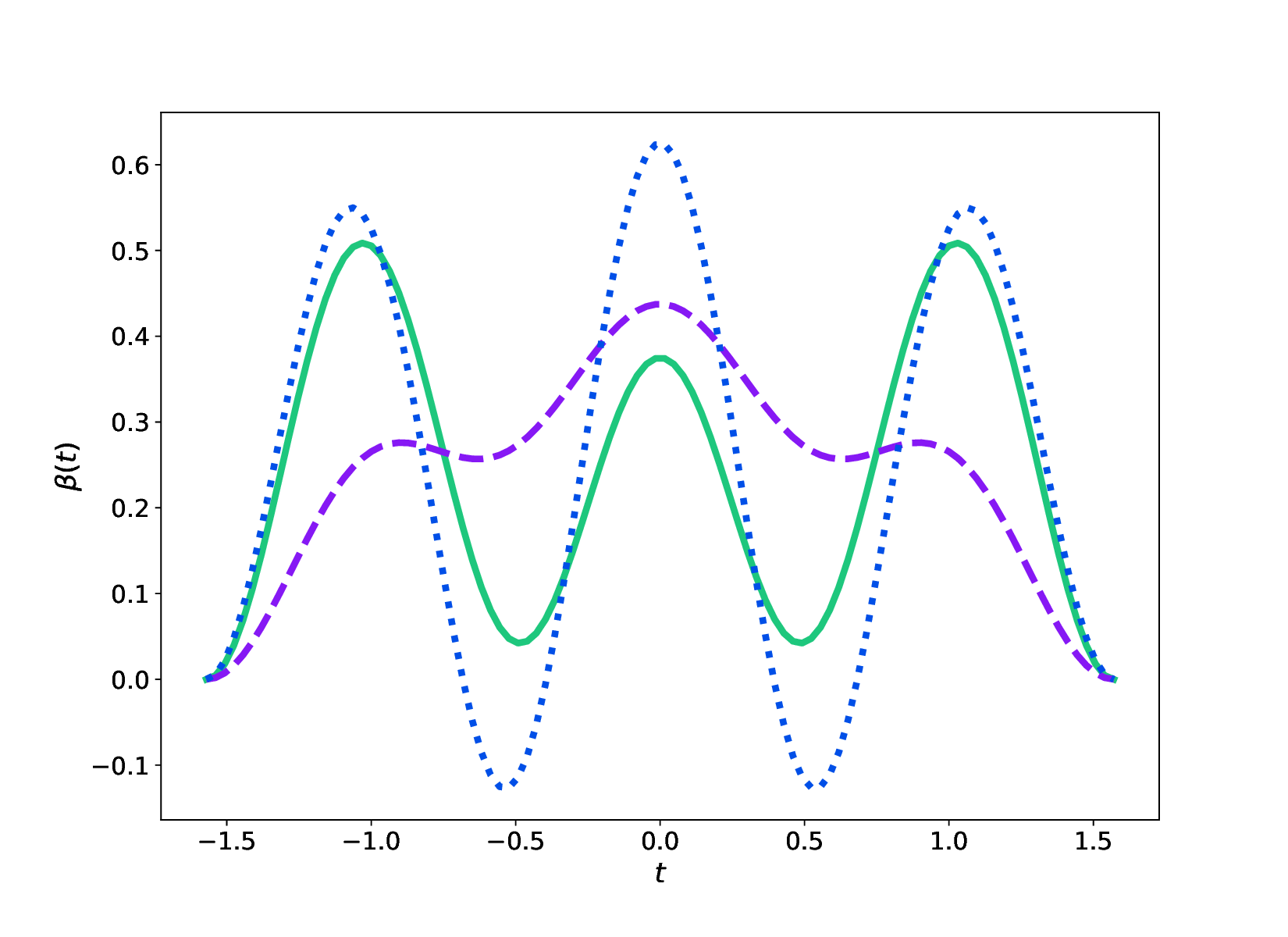}
\caption{The temporal profile of three polyharmonic pulses $\beta(t)$ in the symmetric interval $\left[-\frac{\pi}{2},\frac{\pi}{2}\right]$, with $b=2,c=-3$ (solid line), $b=\frac 9 5, c=-\frac 7 2$ (dashed), and $b=2,c=-5$ (dotted). All three pulses approach zero at the interval's endpoints. However, only the amplitudes depicted with solid and dashed lines can facilitate squeezing transformations of the form \eqref{squeezing}, as $\beta(t)>0$ in both instances. Conversely, the field amplitude depicted with a dotted curve may induce a loop effect on the canonical variables evolution---potentially beneficial for ion traps---where the matrix $h(t)$ would satisfy the condition $|\Sigma|<2$ for complete stability.}
\label{f:pulses}
\end{center}
\end{figure}

As a result, the coefficients of $\theta(t)$ can be obtained straightforwardly by resolving the preceding linear system. Using these coefficients, we can generate pulses $\beta(t)$, which are illustrated through numerical examples in Fig. \ref{f:pulses}. For these examples, we made the choice $\beta\left(-\frac{\pi}{2}\right)=\beta\left(\frac{\pi}{2}\right)=0$.

%In response, the coefficients of $\theta(t)$ become:
%\begin{equation}
%\begin{split}
%a_1 = \frac{(105 b+c+58)b-10}{128 b}, &\quad a_3 = -\frac{(35 b-c-74)b+2}{128 b}, \\
%a_5 = \frac{18-(21 b+c-22)b}{384 b}, &\quad a_7 = \frac{(15 b-c-26)b-6}{384 b},
%\end{split}
%\end{equation}
%the generated pulses $\beta(t)$, based on these coefficients, are illustrated by numerical examples in Fig. \ref{f:pulses}, wherein $\beta\left(-\frac{\pi}{2}\right)=\beta\left(\frac{\pi}{2}\right)=0$ was chosen.

In contrast to the biharmonic field-based method, it is not feasible to generate an equivalent map to the one showcased in Fig. \ref{f:strutt} in this case. Despite this, the polyharmonic pulses, in combination with functions \eqref{exact} and \eqref{theta}, still provide valuable insight into the effects at the end of the symmetric interval of operation. Should $\beta(t)>0$, the magnetic field would induce a squeezing effect on the canonical observables over the concatenated intervals $\left[-\frac{\pi}{2},\frac{\pi}{2}\right]$ and $\left[\frac{\pi}{2},\frac{3\pi}{2}\right]$. Conversely, should $\beta(t)\leq0$, the magnetic field would manifest a stable fuzzy orbit, either closed (loop) or open-ended.

It is important to note that the exact trajectory spread over the interval $\left[-\frac{\pi}{2},\frac{\pi}{2}\right]\cup \left[\frac{\pi}{2},\frac{3\pi}{2}\right]$ remains undetermined at this stage, necessitating a separate computational routine for the integration of \eqref{hdiff} in the asymmetric interval, with $t_0=-\frac{\pi}{2}$. That is to say, the integration of \eqref{hdiff} will be broken down into two subintervals, each defined by a distinct pulse $\beta(t)$. The evolution commences at $-\frac \pi 2$ with the first pulse, continues with the integration of the subsequent subinterval at $\frac \pi 2$ with the second pulse, and concludes at $\frac{3}{2\pi}$, thereby outlining a consistent trajectory.

\begin{figure}[h]
\begin{center}
\includegraphics[width=1.05\textwidth]{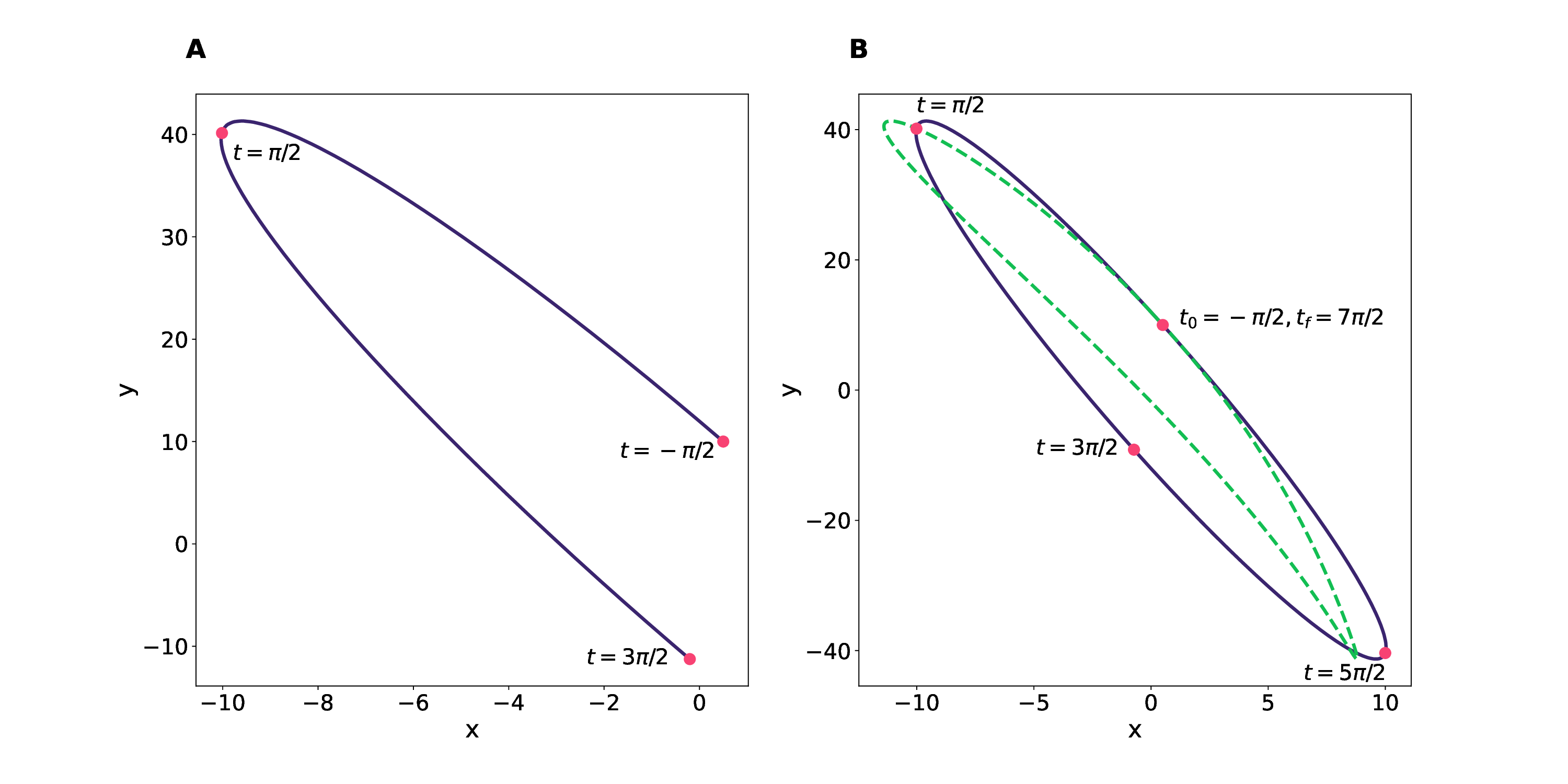}
\caption{Semiclassical trajectories arising from the sequential application of two polyharmonic pulses as described by Equation \eqref{exact}, exerted on a particle with initial state variables $(x,y) = \left(\frac{1}{2}, 10\right)$ and $(p_x, p_y) = (-5, 20)$. {\bf A}. The time evolution begins at $t_0 = -\frac{\pi}{2}$, following a polyharmonic amplitude stipulated by Equation \eqref{exact} and guided by the parameters $b = 2, c = -3$. This phase terminates at $t = \frac{\pi}{2}$, coinciding with the amplitude reaching nullity. At this exact moment, the evolution persists, now steered by the pulse defined by $b = \frac{9}{5}, c = -\frac{7}{2}$, which ceases its impact at $t = \frac{3\pi}{2}$. The resultant operation accomplishes a squeezing transformation with $\lambda_x = -\frac{2}{5}$ and $\lambda_y = -\frac{9}{8}$. {\bf B}. In order to reverse the operation and restore the initial values of the observables, the second pulse must be reapplied, extending from $t = \frac{3\pi}{2}$ to $t = \frac{5\pi}{2}$. Thereafter, the evolution process is once again dominated by the inaugural pulse during the period from $t = \frac{5\pi}{2}$ to $t = \frac{7\pi}{2}$. The dotted line delineates the perturbed evolution engendered by the influence of an external harmonic force $\sin(t)$ acting along the $Ox$ direction. Nonetheless, the original configuration is completely reinstated by inverting the operation.}
\label{f:operations}
\end{center}
\end{figure}

As exemplified in Fig. \ref{f:pulses}, the ensuing control operation is devoid of any abrupt discontinuity, even when it is driven by the concatenation of two differing fields. For instance, the amplitudes $\beta(t)$ represented by solid and dashed curves can facilitate a squeezing transformation as per \eqref{squeezing}. The exact trajectory in the $xOy$ plane for a charged particle with initial conditions $(x,y)=\left(\frac{1}{2}, 10\right)$ and $(p_x,p_y)=(-5,20)$ is plotted in Fig. \ref{f:operations}-A. The canonical position observables undergo a transformation as $X\to\lambda_xX$ and $Y\to\lambda_yY$ in accordance with \eqref{squeezing}. Our numerical calculations reveal that such a particle would be subject to a squeezing transformation with $\lambda_x=-2/5$ and $\lambda_y=-9/8$, indicating that the momenta have been magnified and condensed, respectively.

%	////////////////////////////////////////////////////////////////////
%
%	~
%
%	////////////////////////////////////////////////////////////////////
\subsection{A Non-commutative Geometry Heritage}

Our discussion now delves into the amplification of these effects by external forces, either time-dependent or static. When considering such a scenario, the evolution process must incorporate not only the unperturbed Hamiltonian \eqref{hd} but also the additional external field. Say, for instance, we have an external force aligned in the $Ox$ direction, $\mathbf{F}=(F,0,0)$. This leads to a perturbed Hamiltonian $\widetilde{H}(t)=H_\perp(t)+FX$. Thus, the corresponding evolution operator is defined as $\widetilde{U}(t)=U(t)W(t)$, where $U(t)$ is the unperturbed evolution operator and $W(t)$ represents the perturbative force's influence. This force abides by the differential equation:
\begin{equation}\label{perturbed}
\frac{dW(t)}{dt} = iFX(t)W(t), \quad W(0)=\mathbb{1},
\end{equation}
here, $X(t)$ represents the time-dependent observable driven by the Heisenberg evolution \eqref{heisenberg}. Adding another layer of complexity, because the evolved observable $X(t)$ is linear in the canonical variables, the commutators $[X(t'),X(t)]$ are simplified to mere numbers. 

Invoking the continuous Baker-Campbell-Hausdorff formula\cite{pleban} indicates that \eqref{perturbed} can be solved in the form $e^{iF\int_0^tdt',X(t')}$, subject to a phase factor $e^{i\phi}$, where $\phi$ is real. This reveals that if $t=T$ designates the loop period in the unperturbed evolution operator, any initial coordinate $\bar{X}_j$ of the fuzzy centre \eqref{fuzzy} can be reconstructed from the integral $\int_0^tdt',X(t')$. Thus, $\widetilde{U}(T)=U(T)W(T)=e^{i\phi}e^{iTF\bar{X}_j}$. This suggests that $\bar{X}_j$ remains unaltered by the passing of $W(t)$, namely 
\[
\widetilde{U}^\dagger(T)\bar{X}_j\widetilde{U}(T)=e^{-iTF\bar{X}_j}\bar{X}_je^{iTF\bar{X}_j}=\bar{X}_j.
\]
Conversely, other coordinates of the fuzzy centre might experience a drift
\[
\widetilde{U}^\dagger(T)\bar{X}_{k}\widetilde{U}(T)=\bar{X}_k-iTF[\bar{X}_j,\bar{X}_k],
\]
with $j\neq k$.

This leads us to an interesting conclusion. Any fuzzy point $(\bar{X}_j,\ldots,\bar{X}_k)$ having non-commutative coordinates will break the loop whenever an external force maneouvers it, causing the centre's coordinates to drift transversally relative to the force. On the flip side, if the fuzzy centre's coordinates commutate, $[\bar{X}_j,\bar{X}_k]=0$, the point $(\bar{X}_j,\ldots,\bar{X}_k)$ will return to its initial position at the end of the period $T$.

For illustrative purposes, we have considered the inverted squeezing operation depicted in Fig. \ref{f:operations}-A and introduced a time-dependent perturbation $\sin(t)$ aligned in the $Ox$ direction to deform the operational trajectory (Fig. \ref{f:operations}-B).

\begin{figure}[h]
\begin{center}
\includegraphics[width=\textwidth]{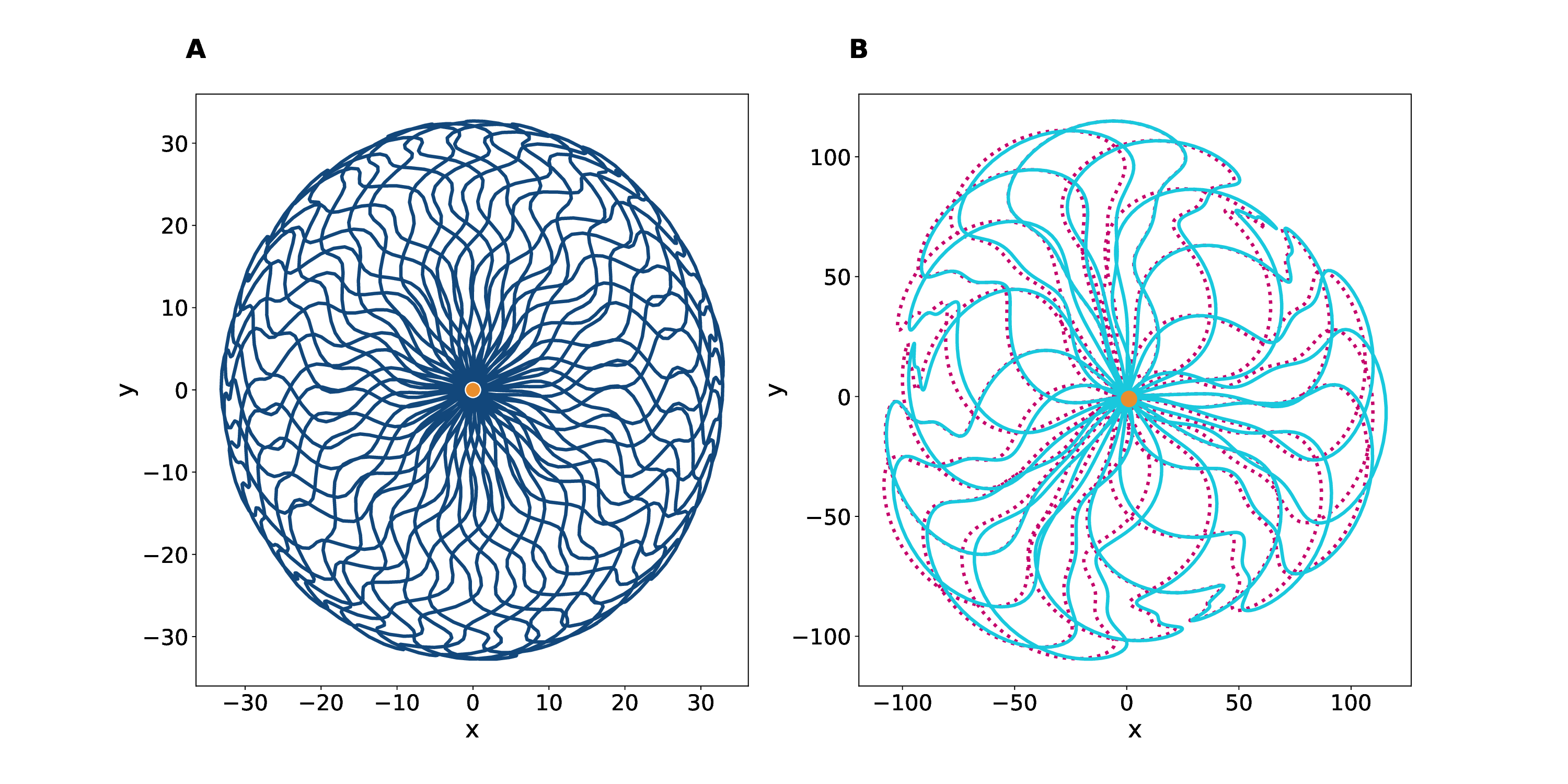}
\caption{{\bf A}. Evolution loops performed via a polyharmonic field amplitude $\beta(t)$ characterised by the parameters $b=2, c=-5$ (see dotted pulse in Fig. \ref{f:pulses}). The charged particle has the initial conditions $(x,y)=(0,0)$ and $(p_x,p_y)=(10,5)$. The operation completes the loop after a period $T=52$. {\bf B}. Distorted evolution loop in terms of the polyharmonic field $\beta(t)$ with parameters $b=\frac{3}{2}, c=-7$ and initial conditions $(x,y)=(1,-1)$, $(p_x,p_y)=(-30,30)$ subject to an external force of magnitude $F=\frac{1}{2}$ in the Ox direction. The solid line represents the unperturbed loop, while the dotted line represents the perturbed loop, closing after a period $T=39$. The perturbed loop exhibits a drifting effect in the opposite direction to the externally applied force. The dot at the centre of both figures represents the initial and final positions.}
\label{f:polyloop}
\end{center}
\end{figure}

As we previously mentioned, the evolution process \eqref{hdiff} initially covers the interval $t\in\left[-\frac{\pi}{2},\frac{\pi}{2}\right]$ with a specific polyharmonic pulse $\beta_1(t)$. The process then continues with a different pulse $\beta_2(t)$ in the interval $t\in\left[\frac{\pi}{2},\frac{3\pi}{2}\right]$. To revert the operation, it necessitates a second application of $\beta_2(t)$ for $t\in\left[\frac{3\pi}{2},\frac{5\pi}{2}\right]$ and a subsequent application of $\beta_1(t)$ for $t\in\left[\frac{5\pi}{2},\frac{7\pi}{2}\right]$. In our particular example, it is noteworthy that the perturbed evolution reverts to its initial state upon the application period's completion. The external force $\sin(t)$ effects a series of transformations \eqref{squeezing} at $t=\frac{3\pi}{2}$, yielding $\lambda_x \approx \frac{3}{7}$ and $\lambda_y \approx -\frac{9}{8}$, unlike the unperturbed case. In this instance, the observable $X$ has been amplified and $Y$ compressed, while simultaneously compressing $P_x$ and amplifying $P_y$.

To underscore the intricacies of this operation, Fig. \ref{f:polyloop} has been provided. We have previously noted that positive polyharmonic pulses \eqref{exact} facilitate the necessary oscillatory motion to effect squeezing operations since the associated matrix $h(t)$ ensures $|\Sigma|>2$. However, a more general class of pulses, potentially positive or negative---such as the amplitude $\beta(t)$ illustrated with a dotted line in Fig. \ref{f:pulses}---and its associated matrix $h(t)$, facilitating stable motion within the region $|\Sigma|<2$, has yet to be explored. As such, this category of pulses $\beta(t)$ offers an alternate route to evolution loop operations, following the same program implemented for biharmonic fields.

Fig. \ref{f:polyloop} provides further illumination on this matter. In panel A, we present a symmetric polyharmonic loop that resists external potential application. Here, resistance signifies absolute stability, much like the examples presented in Fig. \ref{f:bi}. In such a scenario, the loop remains unbroken despite external forces, and any solution to Eq. \eqref{perturbed} will manifest as a phase factor. Conversely, the loop in panel B has been deformed by an external field of constant magnitude (see figure caption for details). It features a fuzzy point \eqref{fuzzy} with non-commutative coordinates. Therefore, its fuzzy centre will experience a transversal drift, which may potentially lead to the loop breaking.

%	////////////////////////////////////////////////////////////////////
%
%	~
%
%	////////////////////////////////////////////////////////////////////
\section{Assortment of Challenges}

The preceding discussion primarily centres around the role of time-varying matrices $h(t)$ in carrying out various quantum control operations. Although the procedure we employed is approximative in nature, such imperfections often add interesting dimensions to the investigation. For instance, in the case of Paul's ion traps\cite{paul}, the electromagnetic signal propagation inside the apparatus is usually ignored due to the small size of the trap. However, adopting a more rigorous stance, even the slight fluctuations in potentials felt on the trap surfaces could contribute to field corrections, beginning from $\frac{1}{c}$ (post-Newtonian) terms in the Einstein-Infeld-Hoffmann (EIH) approximation\cite{infeld}. Here, we briefly address a comparable problem for slowly varying magnetic fields $\beta(t)$ as outlined in \eqref{bi} or \eqref{exact}.

This study is predicated on a homogeneous, time-varying magnetic field $\mathbf{B}(t,\mathbf{x})$ within a cylindrical solenoid. While it does not perfectly adhere to Maxwell's equations, it adheres closely to a series of EIH approximations. In order to understand the magnitude of the resulting errors, we need to compute the exact time-dependent vector potentials for a cylindrical solenoid as given in \eqref{vector}:
\begin{equation}\label{solenoid}
\mathbf{A}(t,\mathbf{x}) = \frac{1}{2}B(t,r){\bf n}\times{\bf x} = \frac{1}{2}B(t,r)\begin{pmatrix}-y\\ x \end{pmatrix},
\end{equation}
the magnetic field $B$ is not just a function of time $t$ but also depends on the radius $r$ of the solenoid's cross section, with $\mathbf{n}$ being a unit vector in the field direction. In order to ensure the relativistic integrity of \eqref{solenoid}, we presume $\square {\bf A}(t,\mathbf{x}) = \frac{4\pi}{c} {\bf J}(t,\mathbf{x}) = 0$, with the symbol $\square$ representing the d'Alembert operator, $\square= \frac{1}{c^2}\frac{\partial^2}{\partial t^2}-\nabla^2$. Upon the application of the Laplacian to the right side of \eqref{solenoid}, we observe that it is equivalent to the operator $D = \frac{\partial^2}{\partial r^2}+\frac{3}{r}\frac{\partial}{\partial r}$ acting on $B(t,r)$. Hence, we can express $\square {\bf A}(t,\mathbf{x})$ as:
\begin{equation}\label{eq:24}
\left[\frac{1}{c^2}\frac{\partial^2}{\partial t^2}-D\right]B(t,r) = 0.
\end{equation}

Assuming that the magnetic field $B(t,r)$ is analytic about $Oz$, we seek a solution of the following form:
\begin{equation}\label{eq:25}
B(t,r) = B_0(t)+B_2(t)r^2+B_4(t)r^4+\cdots,
\end{equation}
where $B_0(t) = B(t)$ signifies the quasi-static homogeneous approximation. Since $Dr^{2n}=4n(n+1)r^{2(n-1)}$, by replacing \eqref{eq:25} into \eqref{eq:24} we get:
\begin{equation}\label{eq:26}
B_{2n}(t) = \frac{1}{4^n n! (n+1)!}\frac{1}{c^{2n}}\frac{\partial^{2n}}{\partial t^{2n}}B(t).
\end{equation}

We now introduce a dimensionless time $\tau = \frac{t}{T}$, where $T$ represents a standard time unit corresponding to laboratory observation, and by expressing \eqref{eq:26} in terms of the time derivatives $\frac{\partial}{\partial\tau}$, we can simplify it to:
\begin{equation}\label{eq:27}
B(t,r) = B(t) + \frac{1}{8}\left(\frac{r}{cT}\right)^2\frac{\partial^2 B(t)}{\partial^2 \tau} + \frac{1}{24}\left(\frac{r}{cT}\right)^4\frac{\partial^4 B(t)}{\partial^4 \tau}+\cdots.
\end{equation}
In this equation, we retain $B(t)$ as a function of actual time $t$ to ensure that all derivatives $\frac{\partial^n}{\partial\tau^n}B(t)$ are expressed in units of magnetic field. This equation intriguingly lacks terms proportional to $\frac{1}{c}$, implying that the field propagation law \eqref{eq:24} is satisfied exclusively by extremely small contributions proportional to $\frac{1}{c^2}$. This implies that the superposition of subtle wave fronts moving towards the solenoid centre form a good approximation to the quasi-static theory.

Addressing how to manifest the initial magnetic step $B(t)$, which kickstarts the entire iterative series mentioned in Eq. \eqref{eq:27}, is a crucial point that needs consideration. In essence, we must determine how to trigger homogeneous surface currents that are time-dependent but spatially constant across $z$.

In a static setting, a stationary current $I$ that circulates around the surface generates the magnetic field within the solenoid, which is, $B = \frac{4\pi}{c}\frac{\Delta I}{\Delta z}$. The conundrum we face, however, is how to produce circulating currents that are time-dependent but homogeneous, and hence independent of $z$, across every section of the surface. One might imagine using a single spiral wire wrapped around the cylindrical surface, with its ends connected to a varying potential difference $\Phi(t)$. However, even a slight change $\delta\Phi(t)$ would propagate as a current pulse along the solenoid, creating a softly varying but $z$-dependent field, as opposed to the desired quasi-static $B(t)$.

An alternative could involve a cylindrical surface made of non-conductive material like glass with radius $a$, uniformly charged with a surface density $\sigma$. The assumption here is that each circular belt with a height of 1cm carries a charge $a\sigma$. This experimental setup is not overly ambitious. Assume a cylinder with a radius $a=20$cm, rotating with an angular velocity $\omega = 1\text{s}^{-1}$ around its symmetry axis aligned with $Oz$ and carrying 1C of charge at every horizontal belt of 1cm. Consequently, a homogeneous magnetic field ${\bf n} B$ of intensity:
\begin{equation}
\label{eq:28}
B = \frac{4\pi}{c}\omega R\sigma\simeq \frac 5 4\text{gauss},
\end{equation}
will be generated inside the cylinder, to a good level of approximation at least. By employing a softly varying angular velocity $\omega\to\omega(t)$, one can indeed create a practically homogeneous magnetic field ${\bf n}B(t)$ that aligns with the quasi-static environment outlined by Eq. \eqref{eq:27}. Yet, does this technique prove effective?

We should note that the actual operation time $\frac{t}{T} \to t$ described in \eqref{variables} could be arbitrarily long (or short), and the fields in \eqref{beta} can be extremely strong (or weak). To provide a realistic estimation of the orders of magnitude needed for our quantum operations, we present Table \eqref{t:1} below.

We also thought it necessary to evaluate the magnitude of the Abraham-Lorentz radiative force. While the conventional force of the variable oscillator trajectory can be expressed as $\mathbf{F}_\text{osc}(t)=m\ddot{\mathbf{x}}(t)$, the proposed radiative force is denoted as $\mathbf{F}_\text{rad}(t) = m\gamma\dddot{\mathbf{x}}(t)$, where $\gamma=\frac{2}{3}\frac{e^2}{mc^3}$ is a characteristic time dependent on the particle. By using the definitions of dimensional quantities in \eqref{variables}, we can compare the magnitudes of conventional and radiative forces for the squeezing operations (refer to the last row in Table \ref{t:1}). The radiative forces are found to be exceptionally small, though they slowly increase as $T$ decreases (higher frequencies).

\begin{table}[h]
\begin{center}\begin{small}
\def\arraystretch{1.8}
\begin{tabular}{|r||l|l|l|}
\hline
$T$ [s]							& $10^{-2}$ 				& 1 					& $10^2$ 						\\\hline
$q$ [cm] 						& 2.5$\times10^{-3}$ 		& 2.5$\times10^{-2}$    & 2.5$\times10^{-1}$ 		\\\hline
$p$ [g cm $\text{s}^{-1}$]		& 4.2$\times10^{-25}$ 		& 4.2$\times10^{-26}$  	& $4.2\times10^{-27}$		\\\hline
$v$ [cm $\text{s}^{-1}$]    	& 2.5$\times10^{-1}$ 	    & 2.5$\times10^{-2}$ 	& 2.5$\times10^{-3}$ 		\\\hline
$B_{\text{max}}$ [gauss] 		& 1.5$\times10^{-2}$  		& 1.5$\times10^{-4}$    & 1.5$\times10^{-6}$		\\\hline
$F_\text{rad}/F_\text{osc}$ 	& 5$\times10^{-25}$     	& 5$\times10^{-27}$     & 5$\times10^{-29}$		\\\hline  
\end{tabular}
\end{small}
\end{center}
\caption{The table presents the necessary physical conditions (in cgs units) for effecting squeezing transformations $\lambda_x\approx-\frac{3}{7}$ and $\lambda_y\approx-\frac{9}{8}$ upon a proton within a cylindrical solenoid. These parameters are linked to the example in Fig. \ref{f:operations}. The quantities of $q$, $p$, and $v$, where $q=x=y$ and $p=p_x=p_y$, are equivalent to the dimensionless values $x'=y'=p'_x=p'_y=1$ in line with Eq. \eqref{variables}. The table notably illustrates the relative increase of the magnetic field strength as the operation time $T$ diminishes. For instance, a control operation lasting $T=10^{-2}$s would require a magnetic field strength half that of Earth's at the equator. In the final row, the average ratio of the Abraham-Lorentz radiative force to the oscillator forces varying over time is shown for differing operation intervals $T$.} 
\label{t:1}
\end{table}

Our preliminary assessments hold true provided that the physical dimensions of the solenoid are achievable. While current laboratory techniques for cooling and trapping ions are effective for probing atomic structure, they often fall short for broader applications. This is largely because the time-dependent oscillator potential can only be generated on a localised scale, as exemplified by a quadrupole trap, whose immediate vicinity from the central axis may merely consist of four metal bars, as seen in certain setups\cite{thompson}.

In our computations, we considered pure states influenced by slowly varying external fields, and did not factor in potential perturbations from traces of matter present across the ion traps or solenoids. We also ignored possible effects of packet reflection or absorption by laboratory walls. These assumptions imply a theoretical apparatus where the dimensions render these considerations negligible. Interestingly, if a particle were to be absorbed by the (surface) walls, we are met with a conceptually challenging problem. This raises the issue of defining a time of arrival for the particle's absorption event, which would necessitate the introduction of a time operator in our quantum mechanical framework. Yet, formulating a time operator that is compatible with the structure of quantum mechanics is a persisting puzzle, dating back to Pauli's objection. While we might intuitively wish to attribute a specific time to the event of absorption, the foundational principles of quantum mechanics make it nontrivial to do so.

We have also observed that in studies employing the Ermakov-Milne invariants, like those by Muga and Berry\cite{muga,berry}, the operations are presumed to be faster than the times we have provided here. In that context, our work takes a slightly different direction: The primary aim is to use a straightforward algebra to derive the exact solutions in \eqref{exact}, while other approaches champion the concept of frictionless driving. This concept involves transitioning states without altering the eigenvalues of certain invariants. But, it begs the question: can these more general variational methods be applied to a $\theta(t)$ function? This is a question that remains to be explored.

%	------------------------------------------------------------/
%
%
%	this is a blank space
%
%
%	------------------------------------------------------------/
\section*{Acknowledgements}

This work is a tribute to Prof. Bogdan Mielnik, whose invaluable insights deeply informed this study. The results presented here manifest our preliminary dialogues.

%	------------------------------------------------------------/
\bibliography{references}
%	------------------------------------------------------------/
\end{document}